\documentclass[12pt,preprint]{aastex}

\shorttitle{Ansae in NGC7009}
\shortauthors{Fern\'{a}ndez et al.}

\begin{document}


\title{Proper motion and kinematics of the ansae in NGC7009}


\author{Rodrigo Fern\'{a}ndez M.\altaffilmark{1}}
\affil{Pontificia Universidad Cat\'{o}lica de Chile, Casilla 306,
Santiago 22, Chile}

\author{Hektor Monteiro, Hugo E. Schwarz}
\affil{Cerro Tololo Inter-American Observatory, NOAO, Casilla 603,
La Serena, Chile.\altaffilmark{2}}
\altaffiltext{1}{CTIO REU-PIA Program student}
\altaffiltext{2}{Cerro Tololo Inter-American Observatory, National Optical
Astronomy Observatory, operated by the Association of Universities for
Research in Astronomy, Inc., under a cooperative agreement with the
National Science Foundation.}



\begin{abstract}
We have measured the proper motion (PM) and kinematics of the ansae in
NGC 7009 using high dispersion echelle spectra and archive narrow band
HST images.  Assuming that the ansae are moving at equal and opposite
velocities from the central star we obtain a system radial velocity of
$-53 \pm 2$ km.s$^{-1}$, the eastern ansa approaching and the western
ansa receding at $v_r=5.3 \pm 1$ km.s$^{-1}$ with respect to this
value.  The PM of the eastern ansa is $28 \pm 8$ mas.yr$^{-1}$, which
with our weighted distance to NGC 7009 of $0.86\pm 0.34$ kpc gives
$V_{exp}=112 \pm 32$ km.s$^{-1}$. The electron temperature and density
in both ansae were determined to be $T_e \sim 9000 \pm 400$ K and $n_e
\sim 2300 \pm 400$ cm$^{-3}$. The dynamic age of the ansae is $\sim
925 \pm 260$ yrs. and the implied PM of the central star is $\mu_{CS}$ = 1
$\pm$ 0.5 mas.yr$^{-1}$. This is in qualitative but not quantitative
agreement with previous work.
\end{abstract}





\section{Introduction}

Most low and intermediate mass stars become White Dwarfs after a brief
phase as Planetary Nebulae (PNe) caused by heavy mass loss on the AGB
\citep{ibr83}. Symmetrical and mildly asymmetrical PNe are
explained by versions of the interacting wind model \citep{kpf78}
but the mechanisms causing the strong asymmetries observed in the
majority of PNe are still unexplained. That both binaries and magnetic
fields probably play a role is now generally accepted. For an overview
of formation mechanisms, models, and asymmetries see \citet{kas00}.

The wealth of different PNe shapes is shown in
\citet{bal87} and \citet{scm92}, and
features have been described in detail by many authors: e.g. point
symmetry \citep{sch93}, bipolarity (e.g. \citealt{csc95}), multipolarity
\citep{sat98}, BRETS \citep{lop97} and Fast, Low
Ionization Emission Regions or FLIERS \citep{bal94} are
all used to indicate the various observed morphologies.

NGC7009 has long been known to have a complex
morphology. \citet{all41} described the outer features as
ansae (handles in Latin), and work by
\citet{rea85}, and \citet{bpi87} show
that these ansae are expanding near the plane of the sky at high
velocities ($\sim 10^2$ km/s) relative to the nucleus.

\citet{lil65} used photographic plate material to
analyze the expansion of NGC7009, including the ansae, for which they
obtained 16 $\pm$ 3 mas.yr$^{-1}$.

In this paper we present the kinematics and proper motion of the ansae
using observational material from the HST and CTIO 4m echelle
spectrograph, and show that the previous results are qualitatively but
not quantitatively correct.

\section{Observations and data reduction}

We extracted two sets of archive HST images with the \emph{``On the
fly''} option. The WFPC2 sets were taken on 1996.04.28 \citep{bal98}
and on 2001.05.11
\citep{pal02} in an H$\alpha$/[NII] filter.

\begin{figure*}[th]
\centering
\includegraphics[scale=0.4]{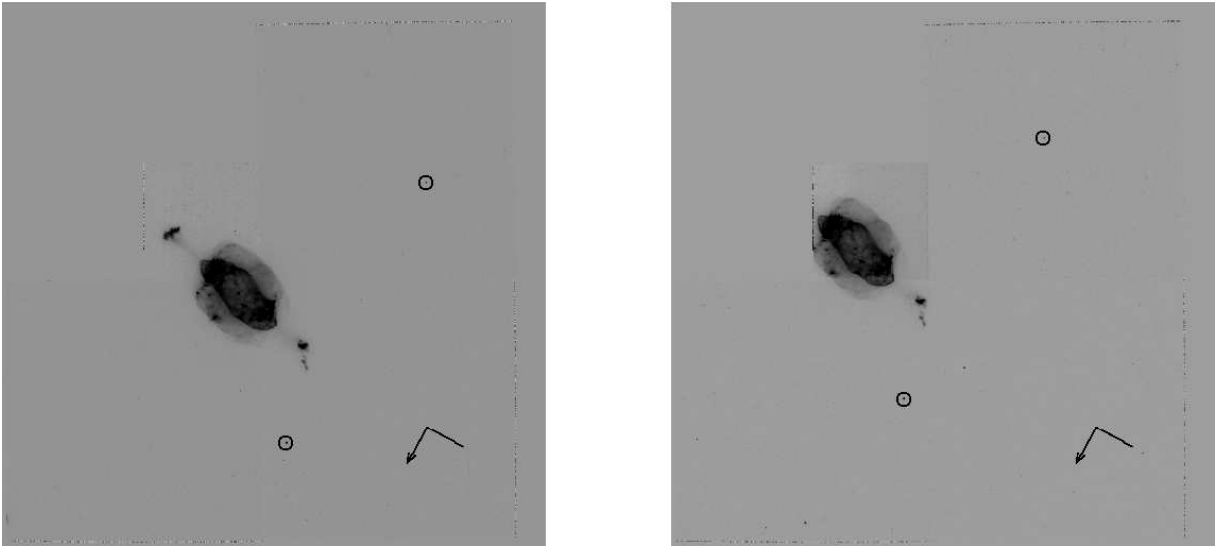}
\caption{Average of the set of HST images taken in 1996 (left) and 
2001 (right). North is down-left and east is down-right, and each
image is 120'' by 120''. There is a shift of 20\," in the east-west
direction between them, which makes the western ansa disappear from
the 2001 images. The two background stars marked with circles were
used together with the CS as reference points for proper motion
calculation.}
\label{fig1}
\end{figure*}

The 2001 images are shifted 20\," east from the 1996
images. As the latter covered the entire nebula, in the former the
west ansa disappears from the field of view. The two images showed in
Figure~\ref{fig1} are an average (for cosmic ray rejection) of
each set.


We took CTIO 4m echelle spectra on 2002.07.29 with a spectral range
between 410--720nm, and a slit of $1.2\arcsec \times 6.6\arcsec$. The
mean seeing was $1.4\arcsec$. Flux calibration to 10\% was done with
\emph{58 Aql} (HR 7596, \citealt{ham92}). The spectra were reduced in
the usual way, using IRAF, and resulted in an rms wavelength error of
0.00037nm, with residuals of $\leq$ 0.00075nm. Velocity resolution was
$3.7$\,km.s$^{-1}$.pix$^{-1}$ or $10.2$\,km.s$^{-1}$ for our slit width at
650nm.

\section{Results}
\subsection{Distance \& proper motion}

The distance to NGC 7009 is uncertain, but we computed the weighted
average of the 14 available \citep{ack92} values
to be $d=0.86\pm 0.34$ kpc.

\begin{figure*}
\begin{center}
\includegraphics[scale=0.3]{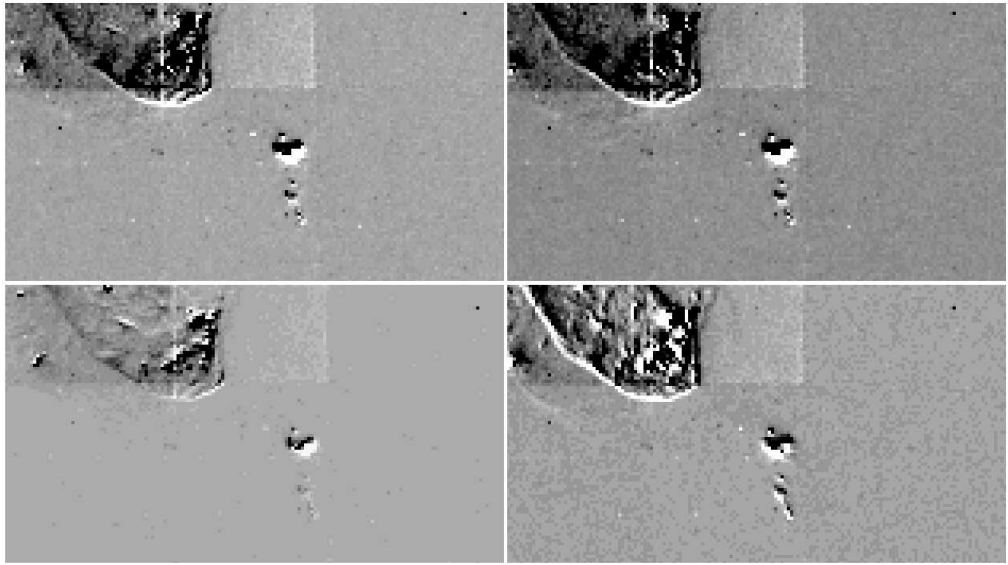}
\end{center}
\caption{Difference between the combined 2001 and 1996 images, normalized
by the average flux of the three reference points chosen: central star
and the two field stars. The panels show differences wrt: central star
(upper--right), field star 1 (lower--left), field star 2 (lower right)
and the adopted least squares fit (upper--left). Each image is 43'' by 24''.}
\label{fig2}
\end{figure*}

The measurement of the central star's position was made with the IRAF
task \emph{imcentroid}. For the measurement of the position of the
ansa, we confined it inside a box of 5'' square and computed the
centroid, using the
\emph{imcentroid} task and measured their displacements with respect
to the central star (CS) and two field stars (shown in
Figure~\ref{fig1}) in two ways: the displacements were measured in all
the images, and an average displacement was obtained for each epoch;
next, the displacements were computed from the averages of the images
for each epoch. The centroid positions given in pixel values were
converted to RA and DEC using the \emph{metric} task, within the
\emph{stsdas} package. This task corrects for the known geometric distortions
between different CCDs. The differences in the results are an order
of magnitude smaller than the errors derived from centroid calculation
and measurements between different chips. The various differential
images are shown in Figure~\ref{fig2}.


\begin{table}[!htb]
\caption{The relative separation between central star and eastern ansa for 
each of the images.}
\begin{tabular}{lcl}
\noalign{\bigskip}
\noalign{\emph{1996 images}}
\noalign{\smallskip}
Dataset &        UT of exp. &   Relative separation (")\\
\noalign{\smallskip}
\hline
\noalign{\smallskip}
u32e0308t &          20:48:16 &                    24.85 $\pm$ 0.04 \\
\noalign{\smallskip}
u32e0309t &          20:56:16 &                    24.84 $\pm$ 0.04 \\
\noalign{\smallskip}
u32e030at &          21:04:16 &                    24.86 $\pm$ 0.04 \\
\noalign{\smallskip}
Average &            20:56:16 &                    24.85 $\pm$ 0.06 \\
\noalign{\smallskip}
\hline
\noalign{\vspace{0.3in}}
\noalign{\emph{2001 images}}
\noalign{\smallskip}
Dataset &       UT of exp. &   Relative separation (")\\
\noalign{\smallskip}
\hline
\noalign{\smallskip}
u5hc6002r &          19:05:14 &                    25.01 $\pm$ 0.04 \\
\noalign{\smallskip}
u5hc6003r &          19:13:14 &                    25.00 $\pm$ 0.04 \\
\noalign{\smallskip}
u5hc6004r &          19:24:14 &                    25.00 $\pm$ 0.04 \\
\noalign{\smallskip}
u5hc6005r &          19:31:14 &                    25.02 $\pm$ 0.04 \\
\noalign{\smallskip}
Average &            19:18:29 &                    25.01 $\pm$ 0.05 \\
\noalign{\smallskip}
\hline
\end{tabular}
\label{relsep}
\end{table}


The displacements between the CS and the ansae are: 24.85$\pm$0.06\arcsec
\& 25.01$\pm$ 0.05\arcsec (E; 1996 \& 2001) and 26.77$\pm$ 0.06\arcsec (W; 1996
only). The detailed measurements with their respective errors are
shown in Table~\ref{relsep}. Note that the adjusted standard
deviations for 1996 and 2001 results are 0.01 and 9.6.10$^{-3}$ respectively.

Even though this is a good determination for the angular displacement
of the eastern ansa, it is also possible that the central star is
moving. Therefore, it is important to have other reference points.
The circles in Figure 1 indicate the position of the two field stars
present on the images that could be measured. The angular
displacements of the ansae in each frame are listed in
Table~\ref{disp}. These measurement are all equal within the errors.


\begin{table}[!htb]
\caption{Angular displacements of the east ansa as measured from 
the combined 1996 and 2001 images using the reference
points specified (Upper star and Lower star refer to
Figure~\ref{fig1}).}

\begin{tabular}{lc}
\noalign{\bigskip}
\hline
Reference Point & Angular displacement (")\\
\noalign{\smallskip}
\hline\hline
\noalign{\smallskip}
Central star & $0.17\pm 0.10$ \\
Upper star & $0.16\pm 0.10$\\
Lower star & $0.10\pm 0.02$\\
Average shift & $0.13\pm 0.04$\\
Least squares shift & $0.14\pm 0.04$\\
\noalign{\smallskip}
\hline
\end{tabular}

\label{disp}
\end{table}


The value we adopted for the displacement was the least squares shift of
all measurements: 0\arcsec.14$\pm$ 0\arcsec.04. The time between images is
158883733s or $\approx$ 5yrs. computed from the start time of the
exposures, yielding a proper motion of the E ansa of 0\arcsec.028 $\pm$
0\arcsec.008 yr$^{-1}$, and the tangential velocity is V$_{exp} = 112 \pm 32$ 
km.s$^{-1}$ for d~=~860pc.

We note that the rim of the inner nebulae also shows differential
motion in more or less the same direction as that of the ansae between
the 1996 and 2001 images. We do not attempt to compute the proper
motion as this rim is likely to be a moving ionization front with a
very different velocity than that of the bulk motion of the gas (weak
R-front, \citet{mell03}). This is unlikely to be the case for the
ansae.

\subsection{Radial Velocities}

We computed the mean observed radial velocities of the ansae from 10
emission lines fitted with a Voigt profile to be $-58.5 \pm
1.9$~km.s$^{-1}$ (E) and $-47.9 \pm 1.6$~km.s$^{-1}$ (W). When using
only the [NII] and H$\alpha$ lines (the HST images were taken in these
lines) we obtain $-58.3 \pm 1.1$~km.s$^{-1}$ (E) and $-47.7
\pm 0.8$~km.s$^{-1}$ (W).  The radial velocity differences between the E and
W ansae are: $\Delta_{H\alpha[NII]} = 10.6 \pm 1.4$~km.s$^{-1}$ and
$\Delta_{All} = 10.6 \pm 2.5$~km.s$^{-1}$. We also separately computed
the mean values of the velocity for the forbidden and for the
permitted lines.  They were also the same to within the errors.  The
observed radial velocities for each detected emission line are listed
in Table~\ref{spectra}, and we also give the heliocentric and LSR
velocities and their average values.


\begin{table}[htb]
\caption{Radial velocities in km.s$^{-1}$ for each emission line identified.} 
\begin{center}
\begin{tabular}{lcccccc}
\hline
 & eastern ansa & & & western ansa & & \\ 
\hline
\noalign{\smallskip}
Line & V$_{obs}$ & V$_{hel}$ & V$_{lsr}$ & V$_{obs}$ & V$_{hel}$ & V$_{lsr}$ \\
\noalign{\smallskip}
\hline
\noalign{\smallskip}
$[$NII$]$  5755  & -59.17 & -55.43 & -45.42 & -47.70 & -43.71 & -33.73 \\
HeI 5876         & -56.42 & -52.68 & -42.67 & -48.51 & -44.56 & -34.54 \\
$[$OI$] $  6364  & -62.23 & -58.49 & -48.48 & -50.44 & -46.49 & -36.47 \\
$[$OI$] $  6300  & -61.19 & -57.45 & -47.44 & -49.05 & -45.10 & -35.08 \\
$[$NII$]$  6548  & -59.33 & -55.59 & -45.58 & -48.34 & -44.39 & -34.37 \\
H$\alpha$  6563  & -56.69 & -52.95 & -42.94 & -48.46 & -44.51 & -34.49 \\
$[$NII$]$  6583  & -57.87 & -54.13 & -44.12 & -46.48 & -42.53 & -32.51 \\
$[$SII$]$  6716  & -57.68 & -53.94 & -43.93 & -45.34 & -41.39 & -31.37 \\
$[$SII$]$  6731  & -56.83 & -53.09 & -43.08 & -45.46 & -41.51 & -31.49 \\
$[$AIII$]$ 7134  & -57.60 & -53.86 & -43.85 & -48.97 & -45.02 & -35.00 \\
\noalign{\smallskip}
{\bf Averages} & -58.5\,$\pm$\,1.9 & -54.76$\pm$\,1.9 & -44.75\,$\pm$\,1.9 & 
-47.9\,$\pm$\,1.6 & -43.92\,$\pm$\,1.6 & -33.91\,$\pm$\,1.6 \\
\noalign{\smallskip}
\hline
\end{tabular}
\end{center}
\label{spectra}
\end{table}

\subsection{Electron temperatures and densities}

The line intensities of [NII] and [SII] were measured by fitting Voigt
profiles to derive the electron temperature and density in both
ansae. The line intensities were corrected for reddening, assuming a
recombination model B with $T = 10^4$ K and $n_e = 10^4$ cm$^{-3}$,
and an extinction coefficient $c_\beta = 0.26$ \citep{pogge96}. The
line intensity ratios together with electron temperatures and
densities derived from them are listed in Table \ref{Tne}. The latter
results were obtained using the formulae from McCall (1984).  The
errors in the ratios were calculated from the S/N in each line.


\begin{table*}[!htb]
\caption{Derived line properties.}
\begin{tabular}{lccl}
\noalign{\bigskip}
\noalign{\emph{Eastern ansa}}
\noalign{\smallskip}
Ratio &                     Observed &       Corrected &        Result\\
\noalign{\smallskip}
\hline
\noalign{\smallskip}
$[$NII$]$(6548+6583)/5755 &      $111 \pm 15$ &       $103 \pm 15$ &        $T_e = 9000 \pm 400$ K\\
\noalign{\smallskip}
$[$SII$]$ 6716/6730 &           $0.67 \pm 0.04$ &    $0.67 \pm 0.04$ &      $n_e = 2600 \pm 500$ $cm^{-3}$\\
\noalign{\smallskip}
\hline
\noalign{\vspace{0.3in}}
\noalign{\emph{Western Ansa}}
\noalign{\smallskip}
Ratio &                     Observed &       Corrected &        Result\\
\noalign{\smallskip}
\hline
\noalign{\smallskip}
$[$NII$]$(6548+6583)/5755 &      $120 \pm 12$ &      $110 \pm 12$ &          $T_e = 8900 \pm 400$ K\\
\noalign{\smallskip}
$[$SII$]$ 6716/6730  &          $0.64 \pm 0.04$ &    $0.64 \pm 0.04$ &        $n_e = 1900 \pm 300$ $cm^{-3}$\\
\noalign{\smallskip}
\hline
\end{tabular}
\label{Tne}
\comment{Line ratios (corrected for reddening), and the results 
obtained from them. $T_e$ and $n_e$ are the electron temperature
and density, respectively.}
\end{table*}

\section{Discussion}

Our radial velocities agree with those of
\citet{rea85}, who determined $\pm\,6.2$~km.s$^{-1}$ using the [OI]6300
line. We get $\pm\,6.1$~km.s$^{-1}$ using the same line.

\citet{lil65} computes 16\,mas.yr$^{-1}$ for the
angular expansion of the ansae; we obtain $28\,\pm\,8$~mas.yr$^{-1}$.
This difference is probably due to the low resolution of the
photographic plates, as Liller comments: ``....the plate scale is such that
photographic grain often competes with seeing as the limiting factor
in image definition".

If the ansae were ejected at the same time from the CS and with equal
\& opposite velocities, we can determine the PM of the CS (relative to
the point of ejection, assumed fixed in space) and the age of the
ansae from our data. We find that the CS has moved 0.96$\pm$\,0.04\,"
since the ejection of the ansae, 920$\pm$\,260 yrs ago. This gives a
proper motion for the CS of 1$\pm$\,0.5 mas.yr$^{-1}$ or 5\,mas between
our two epochs, negligible cf. the motion of the ansae.

In the gas of NGC7009, the sound speed is about 10\,km.s$^{-1}$,
implying that the ansae move at supersonic speeds.

The T$_e$ and n$_e$ we obtain are similar to those found by others:
\citet{bal94} quote a value of T$_e$\,=\,$8100$K \&
n$_e$\,=\,$1000$\,$cm^{-3}$;
\citet{boh94} quote a value of T$_e$\,=\,$9800$K \& n$_e$\,=\,$2300$\,$cm^{-3}$, both for the W ansa.

\section{Conclusions}

We conclude that the ansae in NGC7009 --which is at a distance of
860$\pm$\,340pc-- move outward at the supersonic velocity of
$\pm$\,112\,km.s$^{-1}$ near the plane of the sky
(2.6\,$\pm$\,1\,degrees), have a dynamic age of 920$\pm$\,260 yrs, a
proper motion of 28\,$\pm$\,8\,mas.yr$^{-1}$, and that the CS likely
moves at 1\,$\pm$\,0.5\,mas.yr$^{-1}$.


\end{document}